\documentclass[12pt]{article}

\usepackage[margin=1in]{geometry}
\usepackage{CJKutf8, hyperref, amsmath, amsthm, authblk, graphicx}

\newtheorem{lemma}{Lemma}
\newtheorem{theorem}{Theorem}
\newtheorem{corollary}{Corollary}
\newtheorem{proposition}{Proposition}
\newtheorem{conjecture}{Conjecture}

\theoremstyle{definition}
\newtheorem{definition}{Definition}
\newtheorem{assumption}{Assumption}

\DeclareMathOperator{\erfc}{erfc}
\DeclareMathOperator{\Span}{span}
\DeclareMathOperator{\tr}{tr}

\usepackage[style=trad-unsrt, citestyle=numeric-comp, giveninits=true, doi=false, url=false, eprint=false, isbn=false]{biblatex}
\begin{document}

\title{Universal eigenstate entanglement of chaotic local Hamiltonians}

\begin{CJK}{UTF8}{gbsn}

\author{Yichen Huang (黄溢辰)\thanks{yichuang@mit.edu} \thanks{Present address: Center for Theoretical Physics, Massachusetts Institute of Technology, Cambridge, Massachusetts 02139, USA.}}

\affil{Institute for Quantum Information and Matter, California Institute of Technology, Pasadena, California 91125, USA}

\maketitle

\end{CJK}

\begin{abstract}

In systems governed by ``chaotic'' local Hamiltonians, we conjecture the universality of eigenstate entanglement (defined as the average entanglement entropy of all eigenstates) by proposing an exact formula for its dependence on the subsystem size. This formula is derived from an analytical argument based on a plausible assumption, and is supported by numerical simulations.

\end{abstract}

\section{Introduction}

Entanglement, a concept of quantum information theory, has been widely used in condensed matter and statistical physics to provide insights beyond those obtained via ``conventional'' quantities. For ground states of local Hamiltonians, it characterizes quantum criticality \cite{HLW94, VLRK03, LRV04, CC04, CC09} and topological order \cite{HIZ05, KP06, LW06, LH08, CGW10, HC15}. The scaling of entanglement reflects physical properties (e.g., correlation decay \cite{BH13, BH15, GH16} and dynamical localization \cite{ZPP08, BPM12, Hua17}) and is quantitatively related to the classical simulability of quantum many-body systems \cite{VC06, SWVC08, Osb12, GHLS15, Hua15}.

Besides ground states, it is also important to understand the entanglement of excited eigenstates. Significant progress has been made for a variety of local Hamiltonians \cite{Deu10, SPR12, DLS13, HNO+13, BN13, HM14, KLW15, YCHM15, DKPR16, VHBR17, DLL18, NWFS+18, GG15, HG19}. In many-body localized systems \cite{NH15, AV15, VM16, HZC17}, one expects an ``area law'' \cite{ECP10}, i.e., the eigenstate entanglement between a subsystem and its complement scales as the boundary (area) rather than the volume of the subsystem \cite{BN13, HNO+13, HM14}. In translationally invariant free-fermion systems, the average entanglement entropy of all eigenstates obeys a volume law with a coefficient depending on the subsystem size due to the integrability of the model \cite{VHBR17}.

In this paper we consider ``chaotic'' quantum many-body systems. We are not able to specify the precise meaning of being chaotic, for there is no clear-cut definition of quantum chaos. Intuitively, this class of systems should include non-integrable models in which energy is the only local conserved quantity. For such systems, there are some widely accepted opinions \cite{Deu10, SPR12, DLS13, DKPR16, GG15}:
\begin{enumerate}
\item The entanglement entropy of an eigenstate for a subsystem smaller than half the system size is (to leading order) equal to the thermodynamic entropy of the subsystem at the same energy density.
\item The entanglement entropy of an eigenstate at the mean energy density (of the Hamiltonian) is indistinguishable from that of a random (pure) state.
\item The entanglement entropy of a generic eigenstate is indistinguishable from that of a random state.
\end{enumerate}

We briefly explain the reasoning behind these opinions. The eigenstate thermalization hypothesis (ETH) states that for expectation values of local observables, a single eigenstate resembles a thermal state with the same energy density \cite{Deu91, Sre94, RDO08}. Opinion 1 is a variant of ETH for entropy. Opinion 2 follows from Opinion 1 and the fact that the entanglement entropy of a random state is nearly maximal \cite{Pag93}. Opinion 3 follows from Opinion 2 because a generic eigenstate is at the mean energy density (Lemma 3 in Ref. \cite{HBZ19}).

These opinions concern the scaling of the entanglement entropy only to leading order. A more ambitious goal is to find the exact value of eigenstate entanglement. We conjecture that the average entanglement entropy of all eigenstates is universal (model independent), and propose a formula for its dependence on the subsystem size. This formula is derived from an analytical argument based on an assumption that characterizes the chaoticity of the model. It is also supported by numerical simulations of a non-integrable spin chain.

The formula implies that by taking into account sub-leading corrections not captured in Opinion 3, a generic eigenstate is distinguishable from a random state in the sense of being less entangled. Indeed, this implication can be proved rigorously for any (not necessarily chaotic) local Hamiltonian. The proof also solves an open problem of Keating et al. \cite{KLW15}.

The paper is organized as follows. Section \ref{rand} gives a brief review of random-state entanglement. Section \ref{rig} proves that for any (not necessarily chaotic) local Hamiltonian, the average entanglement entropy of all eigenstates is smaller than that of random states. Sections \ref{chao} and \ref{num} provide an analytical argument and numerical evidence, respectively, for the universality of eigenstate entanglement in chaotic systems. The main text of this paper should be easy to read, for most of the technical details are deferred to Appendices \ref{A} and \ref{B}.

\section{Entanglement of random states} \label{rand}

We begin with a brief review of random-state entanglement. We use the natural logarithm throughout this paper.

\begin{definition} [entanglement entropy]
The entanglement entropy of a bipartite pure state $\rho_{AB}=|\psi\rangle\langle\psi|$ is defined as the von Neumann entropy
\begin{equation}
S(\rho_A)=-\tr(\rho_A\ln\rho_A)
\end{equation}
of the reduced density matrix $\rho_A=\tr_B\rho_{AB}$. It is the Shannon entropy of $\rho_A$'s eigenvalues, which form a probability distribution as $\rho_A\ge0$ (positive semidefinite) and $\tr\rho_A=1$ (normalization).
\end{definition}

\begin{theorem} [conjectured and partially proved by Page \cite{Pag93}; proved in Refs. \cite{FK94, San95, Sen96}] \label{pagethm}
Let $\rho_{AB}$ be a bipartite pure state chosen uniformly at random with respect to the Haar measure. In average,
\begin{equation} \label{page}
S(\rho_A)=\sum_{k=d_B+1}^{d_Ad_B}\frac{1}{k}-\frac{d_A-1}{2d_B}=\ln d_A-\frac{d_A}{2d_B}+\frac{O(1)}{d_Ad_B},
\end{equation}
where $d_A\le d_B$ are the local dimensions of subsystems $A$ and $B$, respectively.
\end{theorem}

Let $\gamma\approx0.5772$ be the Euler-Mascheroni constant. The second step of Eq. (\ref{page}) uses the formula
\begin{equation}
\sum_{k=1}^{d_B}\frac{1}{k}=\ln d_B+\gamma+\frac{1}{2d_B}+O(1/d_B^2).
\end{equation}

\section{Rigorous bounds on eigenstate entanglement} \label{rig}

This section proves a rigorous upper bound on the average entanglement entropy of all eigenstates. The bound holds for any (not necessarily chaotic) local Hamiltonian, and distinguishes the entanglement entropy of a generic eigenstate from that of a random state.

For ease of presentation, consider a chain of $n$ spin-$1/2$'s governed by a local Hamiltonian
\begin{equation} \label{LH}
H=\sum_{i=1}^nH_i,\quad H_i:=H'_i+H'_{i,i+1},
\end{equation}
where $H'_i$ acts only on spin $i$, and $H'_{i,i+1}$ represents the nearest-neighbor interaction between spins $i$ and $i+1$. We use periodic boundary conditions by identifying the indices $i$ and $(i\mod n)$. Suppose $H'_i$ and $H'_{i,i+1}$ are linear combinations of one- and two-local Pauli operators, respectively, so that $\tr H'_i=\tr H'_{i,i+1}=0$ (traceless) and $\tr(H_iH_{i'})=0$ for $i\neq i'$. We assume translational invariance and $\|H_i\|=1$ (unit operator norm). Let $d:=2^n$ and $\{|j\rangle\}_{j=1}^d$ be a complete set of translationally invariant eigenstates of $H$ with corresponding eigenvalues $\{E_j\}$.

\begin{lemma} \label{eigent}
Consider the spin chain as a bipartite quantum system $A\otimes B$. Subsystem $A$ consists of spins $1$,$2$,\ldots,$m$. Assume without loss of generality that $m$ is even and $f:=m/n\le1/2$. Then,
\begin{equation} \label{leq}
S(\rho_{j,A})\le m\ln2-\frac{fE_j^2}{4n},
\end{equation}
where $\rho_{j,A}:=\tr_B|j\rangle\langle j|$ is the reduced density matrix of $|j\rangle$ for $A$.
\end{lemma}

\begin{proof}
See Appendix \ref{A}.
\end{proof}

We are now ready to prove the main result of this section:
\begin{theorem} \label{main1}
In the setting of Lemma \ref{eigent},
\begin{equation} \label{6}
\bar S:=\frac{1}{d}\sum_{j=1}^dS(\rho_{j,A})\le m\ln2-\frac{f\langle H_1^2\rangle}4,
\end{equation}
where $\langle\cdots\rangle:=d^{-1}\tr\cdots$ denotes the expectation value of an operator at infinite temperature.
\end{theorem}

\begin{proof}
Theorem \ref{main1} follows from Lemma \ref{eigent} and the observation that
\begin{equation} \label{spec}
\frac{1}{d}\sum_{j=1}^dE_j^2=\langle H^2\rangle=\sum_{i,i'=1}^n\langle H_iH_{i'}\rangle=\sum_{i=1}^n\langle H_i^2\rangle=n\langle H_1^2\rangle.
\end{equation}
\end{proof}

Recall that Theorem \ref{main1} assumes $\|H_i\|=1$. Without this assumption, (\ref{6}) should be modified to
\begin{equation}
\bar S\le m\ln2-\frac{f\langle H_1^2\rangle}{4\|H_1\|^2}.
\end{equation}

For $2\le m=O(1)$, Theorem \ref{main1} gives the upper bound
\begin{equation}
\bar S\le m\ln2-\Theta(1/n).
\end{equation}
A lower bound can be easily derived from Theorem 1 in Ref. \cite{KLW15}
\begin{equation}
\bar S\ge m\ln2-\Theta(1/n).
\end{equation}
Therefore, both bounds are tight. This answers an open question in Section 6.1 of Ref. \cite{KLW15}.

Without translational invariance (e.g., in weakly disordered systems), a similar result is obtained by averaging over all possible ways of ``cutting out'' a region of length $m$. Here, $\|H_i\|$ may be site dependent but should be $\Theta(1)$ for all $i$.

\begin{corollary} \label{coro}
The average entanglement entropy $\bar{\bar S}$ of a random eigenstate for a random consecutive region of size $m$ is upper bounded by $m\ln2-\Theta(f)$.
\end{corollary}

\begin{proof}
First, we follow the proof of Lemma \ref{eigent}. Without translational invariance, (\ref{c}) remains valid upon replacing $\epsilon_{j,i}$ by $\epsilon_{j,i}/\|H_i\|=\Theta(\epsilon_{j,i})$. By the RMS-AM inequality and Eq. (\ref{spec}), we have
\begin{multline}
\bar{\bar S}\le m\ln2-\frac{\Theta(f)}{4d}\sum_{j=1}^d\sum_{i=1}^{n}\epsilon_{j,i}^2\le m\ln2-\frac{\Theta(f)}{4dn}\sum_{j=1}^dE_j^2=m\ln2-\frac{\Theta(f)}{4n}\sum_{i=1}^n\langle H_i^2\rangle\\
=m\ln2-\Theta(f).
\end{multline}
\end{proof}

It is straightforward to extend all the results of this section to higher spatial dimensions.

\section{Eigenstate entanglement of ``chaotic'' Hamiltonians} \label{chao}

Suppose the Hamiltonian (\ref{LH}) is chaotic in a sense to be made precise below. This section provides an analytical argument for

\begin{conjecture} [universal eigenstate entanglement] \label{main}
Consider the spin chain as a bipartite quantum system $A\otimes B$. Subsystem $A$ consists of spins $1$,$2$,\ldots,$m$. For a fixed constant $f:=m/n\le1/2$, the average entanglement entropy of all eigenstates is
\begin{equation} \label{uni}
\bar S=m\ln2+\frac{\ln(1-f)}2-\frac{2\delta_{f,1/2}}\pi
\end{equation}
in the thermodynamic limit $n\to+\infty$, where $\delta$ is the Kronecker delta.
\end{conjecture}

We split the Hamiltonian (\ref{LH}) into three parts: $H=H_A+H_\partial+H_B$, where $H_{A(B)}$ contains terms acting only on subsystem $A(B)$, and $H_\partial=H'_{m,m+1}+H'_{n,1}$ consists of boundary terms. Let $\{|j\rangle_A\}_{j=1}^{2^m}$ and $\{|k\rangle_B\}_{k=1}^{2^{n-m}}$ be complete sets of eigenstates of $H_A$ and $H_B$ with corresponding eigenvalues $\{\epsilon_j\}$ and $\{\varepsilon_k\}$, respectively. Since $H_A$ and $H_B$ are decoupled from each other, product states $\{|j\rangle_A|k\rangle_B\}$ form a complete set of eigenstates of $H_A+H_B$ with eigenvalues $\{\epsilon_j+\varepsilon_k\}$. Due to the presence of $H_\partial$, a (normalized) eigenstate $|\psi\rangle$ of $H$ with eigenvalue $E$ is a superposition
\begin{equation} \label{superp}
|\psi\rangle=\sum_{j=1}^{2^m}\sum_{k=1}^{2^{n-m}}c_{jk}|j\rangle_A|k\rangle_B.
\end{equation}

The locality of $H_\partial$ implies a strong constraint stating that the population of $|j\rangle_A|k\rangle_B$ is significant only when $\epsilon_j+\varepsilon_k$ is close to $E$.

\begin{lemma} 
There exist constants $c,\Delta>0$ such that
\begin{equation} \label{constr}
\sum_{|\epsilon_j+\varepsilon_k-E|\ge\Lambda}|c_{jk}|^2\le ce^{-\Lambda/\Delta}.
\end{equation}
\end{lemma}

\begin{proof}
This is a direct consequence of Theorem 2.3 in Ref. \cite{AKL16}.
\end{proof}

In chaotic systems, we expect
\begin{assumption} \label{asmp}
The expansion (\ref{superp}) of a generic eigenstate $|\psi\rangle$ is a random superposition subject to the constraint (\ref{constr}).
\end{assumption}

This assumption is consistent with, but goes beyond, the semiclassical approximation Eq. (16) of Ref. \cite{DLL18}.

We now show that Assumption \ref{asmp} implies Conjecture \ref{main}. Consider the following simplified setting. Let $M_j$ be the set of computational basis states with $j$ spins up and $n-j$ spins down, and $U_j\in\mathcal U(|M_j|)=\mathcal U({n\choose j})$ be a Haar-random unitary on $\Span M_j$. Define $M'_j=\{U_j|\phi\rangle:\forall|\phi\rangle\in M_j\}$ so that $M:=\bigcup_{j=0}^nM'_j$ is a complete set of eigenstates of the Hamiltonian
\begin{equation}
H=\sum_{i=1}^n\sigma_i^z.
\end{equation}

The set $M$ captures the essentials of Assumption \ref{asmp}. Every state in $M$ satisfies
\begin{equation}
\sum_{|\epsilon_j+\varepsilon_k-E|\ge 1}|c_{jk}|^2=0,
\end{equation}
which is a hard version of the constraint (\ref{constr}). The random unitary $U_j$ ensures that Eq. (\ref{superp}) is a random superposition. Thus, we establish Conjecture \ref{main} by

\begin{proposition} \label{heu}
The average entanglement entropy of all states in $M$ is given by Eq. (\ref{uni}).
\end{proposition}

\section{Numerics} \label{num}

To provide numerical evidence for Conjecture \ref{main}, consider the spin-$1/2$ Hamiltonian \cite{BCH11, KH13}
\begin{equation} \label{hastings}
H=\sum_{i=1}^nH_i,\quad H_i:=\sigma_i^z\sigma_{i+1}^z+g\sigma_i^x+h\sigma_i^z,
\end{equation}
where $\sigma_i^x,\sigma_i^y,\sigma_i^z$ are the Pauli matrices at site $i(\le n)$, and $\sigma_{n+1}^z:=\sigma_1^z$ (periodic boundary condition). For generic values of $g,h$, this model is non-integrable in the sense of Wigner-Dyson level statistics \cite{BCH11, KH13}. We compute the average entanglement entropy of all eigenstates by exact diagonalization in every symmetry sector.

\begin{figure}
\includegraphics[width=0.5\linewidth]{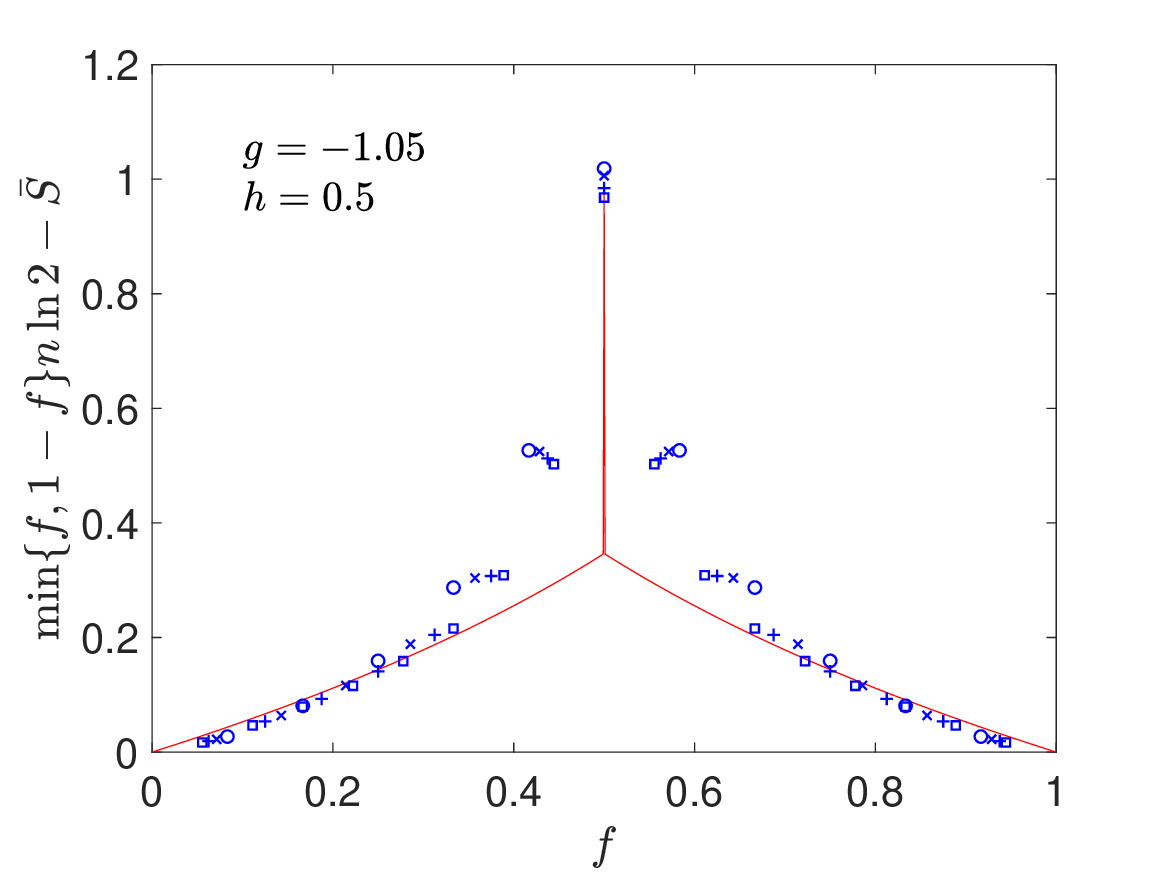}
\includegraphics[width=0.5\linewidth]{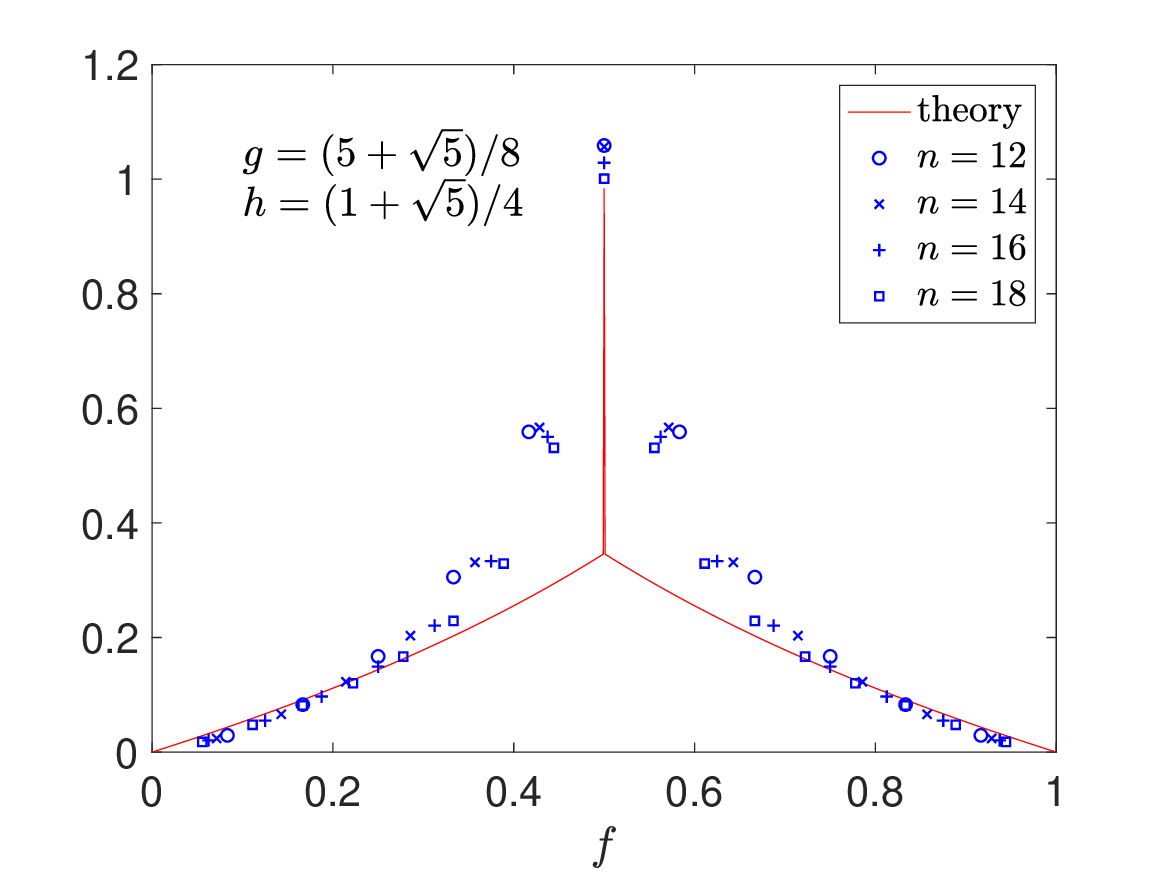}
\caption{Numerical check of Conjecture \ref{main} for two sets of parameters $(g,h)$. The horizontal axes are the fraction $f$ of spins in subsystem $A$. To be aesthetically pleasing, we allow $0<f<1$ so that the plots are mirror symmetric with respect to $f=1/2$. Blue symbols represent corrections obtained by subtracting the average entanglement entropy $\bar S$ of all eigenstates from the leading-order term $\min\{f,1-f\}n\ln2$. Different symbols correspond to different system sizes. Red curves are the theoretical prediction given by Eq. (\ref{uni}).}
\label{plot}
\end{figure}

Figure \ref{plot} shows the numerical results, which semiquantitatively support Conjecture \ref{main}. Noticeable deviations from Eq. (\ref{uni}) are expected due to significant finite-size effects. However, the trend appears to be that the difference between theory and numerics decreases as the system size increases.

\section*{Acknowledgments and notes}

We would like to thank Fernando G.S.L. Brand\~ao and Xiao-Liang Qi for interesting discussions and Anatoly Dymarsky for insightful comments. We acknowledge funding provided by the Institute for Quantum Information and Matter, an NSF Physics Frontiers Center (NSF Grant PHY-1733907). Additional funding support was provided by NSF DMR-1654340.

After this paper appeared on arXiv, we became aware of a simultaneous work \cite{VR17} and a slightly later one \cite{LG17}.

\appendix

\section{Proof of Lemma \ref{eigent}} \label{A}

Let $\epsilon_{j,i}:=\langle j|H_i|j\rangle$ so that $|\epsilon_{j,i}|\le1$. Let $\rho_{j,i}$ be the reduced density matrix of $|j\rangle$ for spins $i$ and $i+1$. Let $I_4$ be the identity matrix of order $4$. Let $\|X\|_1:=\tr\sqrt{X^\dag X}$ be the trace norm. Since $H_i$ is traceless, $|\epsilon_{j,i}|$ provides a lower bound on the deviation of $\rho_{j,i}$ from the maximally mixed state:
\begin{multline}
|\epsilon_{j,i}|=|\tr(\rho_{j,i}H_i)|=|\tr((\rho_{j,i}-I_4/4)H_i)|\le\|\rho_{j,i}-I_4/4\|_1\|H_i\|=\|\rho_{j,i}-I_4/4\|_1\\
=\sum_{k=1}^4|\lambda_k-1/4|,
\end{multline}
where $\lambda_k$'s are the eigenvalues of $\rho_{j,i}$. An upper bound on $S(\rho_{j,i})$ is
\begin{equation}
\max\left\{-\sum_{k=1}^4p_k\ln p_k\right\};\quad\textnormal{s. t.}\quad\sum_{k=1}^4p_k=1,\quad\sum_{k=1}^4|p_k-1/4|\ge|\epsilon_{j,i}|.
\end{equation}
Since the Shannon entropy is Schur concave, it suffices to consider
\begin{itemize}
\item $p_1=p_2=1/4+\epsilon_{j,i}/4$, $p_3=p_4=1/4-\epsilon_{j,i}/4$;
\item (if $\epsilon_{j,i}\ge-1/2$) $p_1=1/4+\epsilon_{j,i}/2$, $p_2=p_3=p_4=1/4-\epsilon_{j,i}/6$;
\item (if $\epsilon_{j,i}\le1/2$) $p_1=1/4-\epsilon_{j,i}/2$, $p_2=p_3=p_4=1/4+\epsilon_{j,i}/6$.
\end{itemize}
For $|\epsilon_{j,i}|\ll1$, by Taylor expansion we can prove
\begin{equation}
S(\rho_{j,i})\le2\ln2-\epsilon_{j,i}^2/2.
\end{equation}
We have checked numerically that this inequality remains valid for any $|\epsilon_{j,i}|\le1$. Therefore,
\begin{equation} \label{c}
S(\rho_{j,A})\le\sum_{k=0}^{m/2-1}S(\rho_{j,2k+1})\le m\ln2-\frac{1}{2}\sum_{k=0}^{m/2-1}\epsilon_{j,2k+1}^2
\end{equation}
due to the subadditivity \cite{AL70} of the von Neumann entropy. We complete the proof using $\epsilon_{j,i}=E_j/n$.

\section{Proof of Proposition \ref{heu}} \label{B}

Assume without loss of generality that $n$ is even. Let $L_j$ ($R_j$) be the set of computational basis states of subsystem $A$ ($B$) with $j$ spins up and $m-j$ ($n-m-j$) spins down so that
\begin{equation}
|L_j|={m\choose j},\quad|R_j|={n-m\choose j},\quad\textnormal{and}\quad M_j=\bigcup_{k=\max\{0,m-n+j\}}^{\min\{m,j\}}L_k\times R_{j-k}.
\end{equation}
Thus, any (normalized) state $|\psi\rangle$ in $M'_j$ can be decomposed as 
\begin{equation}
|\psi\rangle=\sum_{k=\max\{0,m-n+j\}}^{\min\{m,j\}}c_k|\phi_k\rangle,
\end{equation}
where $|\phi_k\rangle$ is a normalized state in $\Span L_k\otimes\Span R_{j-k}$. Let $\rho_A$ and $\sigma_{k,A}$ be the reduced density matrices of $|\psi\rangle$ and $|\phi_k\rangle$ for $A$, respectively. It is easy to see
\begin{equation} \label{rhoa}
\rho_A=\bigoplus_{k=\max\{0,m-n+j\}}^{\min\{m,j\}}|c_k|^2\sigma_{k,A}\implies S(\rho_A)=\sum_{k=\max\{0,m-n+j\}}^{\min\{m,j\}}|c_k|^2S(\sigma_{k,A})-|c_k|^2\ln|c_k|^2.
\end{equation}
Since $|\psi\rangle$ is a random state in $\Span M_j$, each $|\phi_k\rangle$ is a (Haar-)random state in $\Span L_k\otimes\Span R_{j-k}$. Theorem \ref{pagethm} implies that in average,
\begin{equation}
S(\sigma_{k,A})=\ln\min\{|L_k|,|R_{j-k}|\}-\frac{\min\{|L_k|,|R_{j-k}|\}}{2\max\{|L_k|,|R_{j-k}|\}}.
\end{equation}
In average, the population $|c_k|^2$ is proportional to the dimension of $\Span L_k\otimes\Span R_{j-k}$:
\begin{equation} \label{ck}
|c_k|^2=|L_k||R_{j-k}|/|M_j|.
\end{equation}
The deviation of $|c_k|^2$ (from the mean) for a typical state $|\psi\rangle\in\Span M_j$ is exponentially small. In the thermodynamic limit, $j,k$ can be promoted to continuous real variables so that $|M_j|,|L_k|$ follow normal distributions with means $n/2,fn/2$ and variances $n/4,fn/4$, respectively. Let
\begin{equation}
J:=j/\sqrt n-\sqrt n/2,\quad K:=k/\sqrt n-f\sqrt n/2.
\end{equation}
We have
\begin{gather}
|L_k|=\sqrt2d^fe^{-2K^2/f}/\sqrt{f\pi n},\quad|R_{j-k}|=\sqrt2d^{1-f}e^{-2(J-K)^2/(1-f)}/\sqrt{(1-f)\pi n},\\
|M_j|=\sqrt2de^{-2J^2}/\sqrt{\pi n},\quad|c_k|^2=\sqrt2e^{2J^2-2K^2/f-2(J-K)^2/(1-f)}/\sqrt{f(1-f)\pi n}.\label{ckexp}
\end{gather}

For any fixed constant $f<1/2$, it is almost always the case that $|L_k|\ll|R_{j-k}|$. Hence,
\begin{equation} \label{ska}
S(\sigma_{k,A})=\left(m+\frac12\right)\ln2-\frac{\ln(f\pi n)}2-\frac{2K^2}f.
\end{equation}
Substituting Eqs. (\ref{ckexp}) and (\ref{ska}) into Eq. (\ref{rhoa}),
\begin{align}
&S_j:=S(\rho_A)=\int_{-\infty}^{+\infty}|c_k|^2(S(\sigma_{k,A})-\ln|c_k|^2)\,\mathrm dk\nonumber\\
&=\left(m+\frac12\right)\ln2-\frac{\ln(f\pi n)}2+\frac{f-4fJ^2-1}2+\frac{1+\ln(f-f^2)+\ln(\pi n/2)}2\nonumber\\
&=m\ln2+\frac{f(1-4J^2)}2+\frac{\ln(1-f)}2.
\end{align}
Averaging over all states in $M$,
\begin{equation} \label{flh}
\bar S=\frac1d\sum_{j=0}^n|M_j|S_j\approx m\ln2+\frac{\ln(1-f)}{2}+\frac{f}{\sqrt{2\pi}}\int_{-\infty}^{+\infty}e^{-2J^2}(1-4J^2)\,\mathrm dJ=m\ln2+\frac{\ln(1-f)}{2}.
\end{equation}

For $f=1/2$, we first assume that $j\le n/2$ and $k\le j/2$ (i.e., $J\le0$ and $K\le J/2$) so that $|L_k|\le|R_{j-k}|$. Hence,
\begin{equation} \label{skae}
S(\sigma_{k,A})=\left(\frac{n}2+1\right)\ln2-\frac{\ln(\pi n)}2-4K^2-\frac{e^{4J^2-8JK}}2.
\end{equation}
Let
\begin{equation}
\erfc x:=\frac{2}{\sqrt\pi}\int_x^{+\infty}e^{-t^2}\,\mathrm dt
\end{equation}
be the complementary error function. Substituting Eqs. (\ref{ckexp}) and (\ref{skae}) into Eq. (\ref{rhoa}),
\begin{align} \label{sj}
&S_j:=S(\rho_A)=2\int_{-\infty}^{j/2}|c_k|^2S(\sigma_{k,A})\,\mathrm dk-\int_{-\infty}^{+\infty}|c_k|^2\ln|c_k|^2\,\mathrm dk\nonumber\\
&=\left(\frac{n}{2}+1\right)\ln2-\frac{\ln(\pi n)}{2}-\frac{1}{4}+J\sqrt{\frac{2}{\pi}}-J^2-\frac{e^{2J^2}\erfc(-\sqrt2J)}{2}+\frac{1+\ln(\pi n/8)}{2}\nonumber\\
&=\frac{n-1}{2}\ln2+\frac{1}{4}+J\sqrt{\frac{2}{\pi}}-J^2-\frac{e^{2J^2}\erfc(-\sqrt2J)}{2}.
\end{align}
This is the average entanglement entropy of a random state in $\Span M_j$ for $j\le n/2$. For $j>n/2$, Eq. (\ref{sj}) remains valid upon replacing $J$ by $-J$. Averaging over all states in $M$,
\begin{multline} \label{fh}
\bar S=\frac{1}{d}\sum_{j=0}^n|M_j|S_j\approx\frac{n-1}{2}\ln2+\sqrt\frac{8}{\pi}\int_{-\infty}^0e^{-2J^2}\left(\frac{1}{4}+J\sqrt{\frac{2}{\pi}}-J^2-\frac{e^{2J^2}\erfc(-\sqrt2J)}{2}\right)\,\mathrm dJ\\
=\frac{(n-1)\ln2}2-\frac2\pi.
\end{multline}
Equation (\ref{uni}) follows from Eqs. (\ref{flh}) and (\ref{fh}).

\printbibliography

\end{document}